\documentclass[twoside,twocolumn,9pt]{article}
\usepackage{extsizes}
\usepackage[super,sort&compress,comma]{natbib} 
\usepackage[version=3]{mhchem}
\usepackage[left=1.5cm, right=1.5cm, top=1.785cm, bottom=2.0cm]{geometry}
\usepackage{balance}
\usepackage{mathptmx}
\usepackage{amsmath}
\usepackage{physics}
\usepackage{amssymb}
\usepackage{sectsty}
\usepackage{graphicx} 
\usepackage{lastpage}
\usepackage[format=plain,justification=justified,singlelinecheck=false,font={stretch=1.125,small,sf},labelfont=bf,labelsep=space]{caption}
\usepackage{float}
\usepackage{fancyhdr}
\usepackage{fnpos}
\usepackage[english]{babel}
\usepackage{array}
\usepackage{droidsans}
\usepackage{charter}
\usepackage[T1]{fontenc}
\usepackage[usenames,dvipsnames]{xcolor}
\usepackage{setspace}
\usepackage[compact]{titlesec}
\usepackage{hyperref}

\usepackage{epstopdf}

\definecolor{cream}{RGB}{222,217,201}

\usepackage{xr}
\makeatletter
\newcommand*{\addFileDependency}[1]{
  \typeout{(#1)}
  \@addtofilelist{#1}
  \IfFileExists{#1}{}{\typeout{No file #1.}}
}
\makeatother
\newcommand*{\myexternaldocument}[1]{%
    \externaldocument{#1}%
    \addFileDependency{#1.tex}%
    \addFileDependency{#1.aux}%
}
\myexternaldocument{supplementary}

\begin{document}

\pagestyle{fancy}
\thispagestyle{plain}
\fancypagestyle{plain}{
\renewcommand{\headrulewidth}{0pt}
}

\makeFNbottom
\makeatletter
\renewcommand\LARGE{\@setfontsize\LARGE{15pt}{17}}
\renewcommand\Large{\@setfontsize\Large{12pt}{14}}
\renewcommand\large{\@setfontsize\large{10pt}{12}}
\renewcommand\footnotesize{\@setfontsize\footnotesize{7pt}{10}}
\makeatother

\renewcommand{\thefootnote}{\fnsymbol{footnote}}
\renewcommand\footnoterule{\vspace*{1pt}%
\color{cream}\hrule width 3.5in height 0.4pt \color{black}\vspace*{5pt}} 
\setcounter{secnumdepth}{5}

\makeatletter 
\renewcommand\@biblabel[1]{#1}            
\renewcommand\@makefntext[1]%
{\noindent\makebox[0pt][r]{\@thefnmark\,}#1}
\makeatother 
\renewcommand{\figurename}{\small{Fig.}~}
\sectionfont{\sffamily\Large}
\subsectionfont{\normalsize}
\subsubsectionfont{\bf}
\setstretch{1.125} 
\setlength{\skip\footins}{0.8cm}
\setlength{\footnotesep}{0.25cm}
\setlength{\jot}{10pt}
\titlespacing*{\section}{0pt}{4pt}{4pt}
\titlespacing*{\subsection}{0pt}{15pt}{1pt}

\fancyfoot{}
\fancyfoot[LO,RE]{\vspace{-7.1pt}\includegraphics[height=9pt]{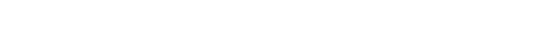}}
\fancyfoot[CO]{\vspace{-7.1pt}\hspace{13.2cm}\includegraphics{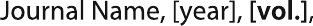}}
\fancyfoot[CE]{\vspace{-7.2pt}\hspace{-14.2cm}\includegraphics{head_foot/RF}}
\fancyfoot[RO]{\footnotesize{\sffamily{1--\pageref{LastPage} ~\textbar  \hspace{2pt}\thepage}}}
\fancyfoot[LE]{\footnotesize{\sffamily{\thepage~\textbar\hspace{3.45cm} 1--\pageref{LastPage}}}}
\fancyhead{}
\renewcommand{\headrulewidth}{0pt} 
\renewcommand{\footrulewidth}{0pt}
\setlength{\arrayrulewidth}{1pt}
\setlength{\columnsep}{6.5mm}
\setlength\bibsep{1pt}

\makeatletter 
\newlength{\figrulesep} 
\setlength{\figrulesep}{0.5\textfloatsep} 

\newcommand{\topfigrule}{\vspace*{-1pt}%
\noindent{\color{cream}\rule[-\figrulesep]{\columnwidth}{1.5pt}} }

\newcommand{\botfigrule}{\vspace*{-2pt}%
\noindent{\color{cream}\rule[\figrulesep]{\columnwidth}{1.5pt}} }

\newcommand{\dblfigrule}{\vspace*{-1pt}%
\noindent{\color{cream}\rule[-\figrulesep]{\textwidth}{1.5pt}} }

\makeatother

\twocolumn[
  \begin{@twocolumnfalse}
{\includegraphics[height=30pt]{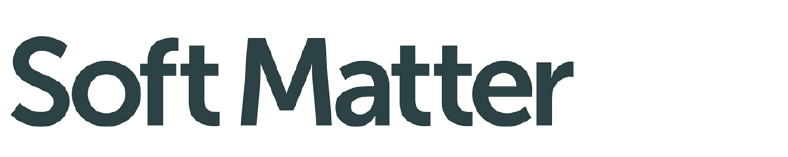}\hfill\raisebox{0pt}[0pt][0pt]{\includegraphics[height=55pt]{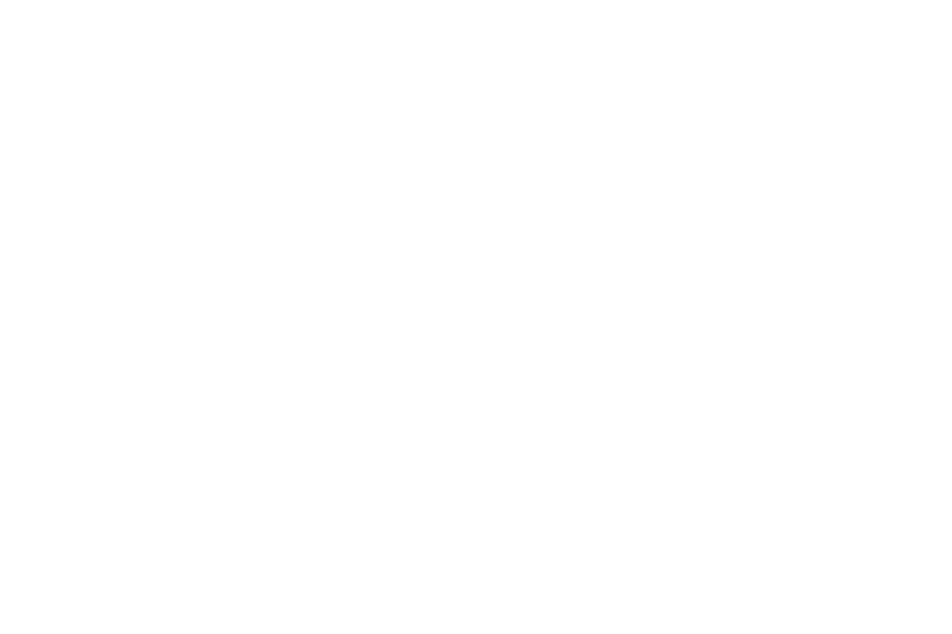}}\\[1ex]
\includegraphics[width=18.5cm]{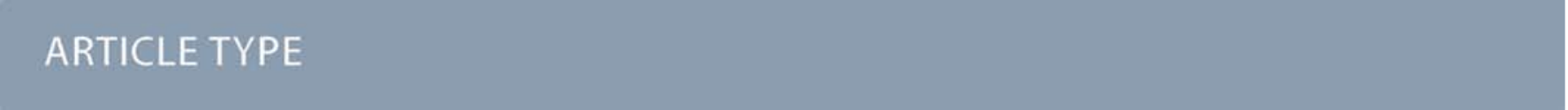}}\par
\vspace{1em}
\sffamily
\begin{tabular}{m{4.5cm} p{13.5cm} }

\includegraphics{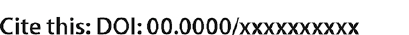} & \noindent\LARGE{\textbf{Packing and emergence of ordering of rods in a spherical monolayer$^\dag$}} \\
\vspace{0.3cm} & \vspace{0.3cm} \\

 & \noindent\large{Dharanish Rajendra\textit{$^{a}$}, Jaydeep Mandal\textit{$^{a}$}, Yashodhan Hatwalne\textit{$^{b}$} and Prabal K. Maiti$^{\ast}$\textit{$^{a}$}} \\

\includegraphics{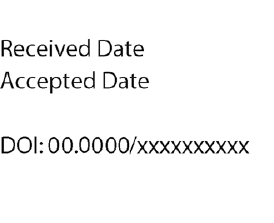} &
\noindent\normalsize{Spatially ordered systems confined to surfaces such as spheres exhibit interesting topological structures because of curvature induced frustration in orientational as well as translational order. The study of these structures is important for investigating the interplay between geometry, topology, and elasticity, and for their potential applications in materials science.
In this work we numerically simulate a spherical monolayer of soft repulsive spherocylinders (SRS) and study the packing of rods and their ordering transition as a function of the packing fraction. In the  model that we study, centers of mass of the spherocylinders (situated at their geometric centers) are constrained to move on a spherical surface. The spherocylinders are free to rotate about any axis 
that passes through their respective centers of mass. 
We show that at relatively lower packing fractions, there is a continuous transition from a disordered fluid to a novel, orientationally ordered, spherical fluid monolayer as the packing fractions is increased. This monolayer of orientationally ordered SRS particles resembles a hedgehog --- long axes of the SRS particles are aligned along the local normal to the sphere. At higher packing fractions, system undergoes transition to the solid phase, which is riddled with topological point defects (disclinations) and grain boundaries that divide the whole surface into several domains. } \\

\end{tabular}

 \end{@twocolumnfalse} \vspace{0.6cm}

  ]

\renewcommand*\rmdefault{bch}\normalfont\upshape
\rmfamily
\section*{}
\vspace{-1cm}

\footnotetext{\textit{$^{a}$~Centre for Condensed Matter Theory, Department of Physics, Indian Institute of Science, Bengaluru 560012, India}}
\footnotetext{\textit{$^{b}$~Raman Research Institute, Bengaluru 560012, India}}
\footnotetext{$^{\ast}$~\textit{Corresponding author. Email: maiti@iisc.ac.in}}

\footnotetext{\dag~Electronic Supplementary Information (ESI) available: [details of any supplementary information available should be included here]. See DOI: 10.1039/cXsm00000x/}



\section{Introduction}
\begin{figure*}
    \centering
    \includegraphics[width=15cm]{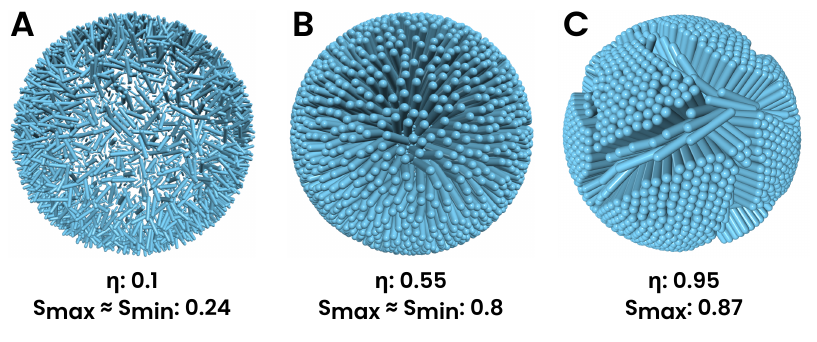}
    \caption{Different phases seen in the system: (A) disordered phase at low packing fraction, (B) two dimensional liquid crystal phase at medium packing fractions, where particles are directed radially outwards, (C) solid phase that occurs at high packing fractions and has multiple domains of high ordering separated by defect lines of low ordering. The ordering in the fluid phase increases with increase in packing fraction. All simulations done for $A=5$. $\eta=a\rho$ is the packing fraction ($a,\rho$ are the cross-sectional area of spherocylinders and surface density, respectively) and $S$ is the nematic order parameter.}
    \label{fig:phases}
\end{figure*}

The statistical mechanics of rod-like particles has been an important problem ever since Onsager developed the theory for the (three dimensional) isotropic-nematic liquid crystalline (IN) phase transition in a system of hard rods \cite{onsager1949effects}. Since then, a number of studies have investigated the different phases and phase transitions for hard and soft rods. \cite{bolhuis_frenkel,cuetos_haya,bates_frenkel,allen,cuetos2003liquid,martinez2007stability,marechal2011phase,van1995transverse,earl2001computer}. Bolhuis and Frenkel \cite{bolhuis_frenkel} have extensively studied the phases of hard rods in bulk and showed that the phases and phase boundaries vary depending on the shape anisotropy $A=L/D$ ($D$ and $L$ are the diameter and core length of the rod, respectively). Cuetos and Martinez-Haya \cite{cuetos_haya} have studied the effect of temperature on the phase diagram by using the mapping equation between soft to hard rods and observed triple points between different phases. Bates and Frenkel \cite{bates_frenkel} have simulated hard rods on 2D plane, and showed that for $A \geq 6$, there is a nematic phase with algebraically decaying orientational correlation, whereas for small shape anisotropies, there is an isotropic phase with strong local positional and orientational correlation. External fields can also introduce novel phases in systems of soft polarizable spherocylinders \cite{glaser}. 
Furthermore, Dussi et. al \cite{dijkstra1} have simulated different single component systems with particles of different shapes, and showed that depending on the system size, a prolate columnar phase appears in the system. But this columnar phase is mechanically unstable as the system size is increased. This phenomena is quite general in the sense that it is observed for all the different particle shapes. The study of phases is not only limited to single component systems of hard and soft rods. Experimental and simulation studies have also been carried out for binary mixture of particles \cite{adams_rod_sphere_expt,dijkstra3,dijkstra2},\cite{dussi2016entropy,dussi2018hard}. In general, the study of Liquid Crystals are quite extensive \cite{de1993physics,mcgrother1996re,maiti2002induced,lansac2003phase,patti2012brownian}.
In the recent years, the study of two dimensional nematic order on curved surfaces (such as spheres) has gained impetus because of possible experimental realizations of such systems \cite{colloidosome1,colloidosome2,microfluidics1}.
Curvature driven dynamics also play an important role in different biological processes \cite{bio2,bio1,janssen_kaiser_lowen} as well as in different properties of colloidal systems \cite{law2018nucleation,law2020phase}\cite{akram2020chiral}. 
Three dimensional uniaxial nematics are orientationally ordered fluids, and can be characterized by the three component unit director $\vu{n} = ( n_{x}(x,y,z), n_{y}(x,y,z), n_{z}(x,y,z) ) $, whereas the unit director of two dimensional nematics in a plane has two components: $\vu{n} = ( n_{x}(x,y), n_{y}(x,y) ) $. On a curved surface such as a sphere, the two dimensional nematic director lies in the local tangent plane to the sphere. However, any such vector or director field on the  sphere is frustrated because of the intrinsic (Gaussian) curvature of the sphere. As is well known, a hairy ball cannot be combed flat without creating at least one hair whorl, a single vortex, or vortices of total strength 2 \cite{poincare1885courbes,Brouwer1912}. In condensed matter physics these topological point defects are called disclinations \cite{de1993physics}. Surrounding the disclination point (eye of the vortex), orientational deformations are very large, destroying the orientational order. Disclinations are characterized by their index, and have ``molten'' core regions of finite extent encompassing the disclination point.  
Because of rich variety of configurations shown by such systems, various numerical studies have also been carried out for analyzing the structures and defects. Lubensky and Prost \cite{lubensky} have theorized that director configuration of nematics on spherical surfaces would have four +1/2 disclinations, which has been verified in the numerical study of Bates \cite{Bates}. 
Interestingly, the arrangement of the defect configuration for nematic liquid crystal on spherical surfaces is observed to alter with elastic anisotropy \cite{dhakal,shin}. The change in elastic anisotropy can be realized by the change in temperature of the system \cite{defect_expt1} and other system environmental conditions.
Along with the defects, numerical studies have also revealed various textures (director fields) for systems in which the rods lie tangent to spherical surface \cite{smallenburg,exotic_structure}.

Disclination cores on spherical particles such as micron-sized colloidal particles coated with liquid crystals can be functionalized to create ``superatoms'' with directional bonds \cite{devries}. This opened up new possibilities of such self-assembly of superatoms by linking across functionalized groups (including biomolecules such as DNA) and the development of atomic chemistry at micron scales. Thin nematic shells consisting of a nematic drop containing a smaller aqueous drop have been obtained in double emulsions \cite{thick_nematic_shell}. These can be engineered to imitate $sp,\ sp^{2}$, and $sp^{3}$ geometries of carbon bonds \cite{nelson_sp3}. Deformable vesicles with orientational order can form facets. These fascinating properties have led to rapid advances in theoretical, and experimental studies \cite{nematic_expt1,nematic_expt2}.
In recent years a new branch of colloidal science called ``topological colloids'' has emerged. When introduced into a nematic liquid crystal, topological colloids induce three dimensional director fields and topological defects dictated by colloidal topology. This lays the groundwork for new application of colloids, such as topological memory devices etc., and the experimental study of low dimensional topology \cite{poulin1997novel,musevic2006two,liu2013nematic,senyuk2013topological,nych2013assembly}.

In this work, we focus on the phases, structural transitions between them, and on topological defects in a spherical monolayer of SRS particles. The rods lie within a spherical shell of inner and outer radii $(R-(L+D)/2)$ and $(R+(L+D)/2)$, respectively, where $R$ is the radius of the sphere on which the center of masses of the rods are constrained to lie, and $D$ and $L$ are the diameter and core length of the rod.

At low packing fraction $\left(\eta \lesssim 0.35\right)$(see section \ref{section_model} for definition of packing fraction $\eta$), the system is almost completely disordered with nematic and radial order parameters (see Section \ref{section_model} Eqn. \ref{eqn:nematic-order-parameter} and \ref{eqn:radial_order_parameter} for definition) close to zero (Fig. \ref{fig:phases}A). At medium densities ($\eta \sim 0.35{-}0.65 $), it adopts an orientationally ordered configuration with the rods all aligned with the local radial direction (Fig. \ref{fig:phases}B). In this phase the nematic and radial order parameters take values up to 0.8 and 1 respectively. This phase does not have positional ordering, therefore, we characterize it as radially oriented, two dimensional liquid crystalline phase. The change in ordering from the disordered phase to the liquid crystal phase (quantified by the nematic and radial order parameters) is completely smooth.
We note that in contrast to two dimensional nematics, the liquid crystalline phase described above has a three component director on a two dimensional spherical surface. Moreover, the ground state configuration of this phase is disclination free, as the hairy ball theorem is not applicable to it --- the director is everywhere normal to the spherical surface. The orientationally ordered sphere itself is the core of a surface (two dimensional) topological defect called a hedgehog of index 2 \cite{chaikin1995principles}. Spheres with this structure are called hedgehog particles \cite{hedgehog}.  

The solid phase occurs at high packing fractions ($\eta \gtrsim 0.65$) (Fig. \ref{fig:phases}C) and shows high degree of positional and orientational ordering. However, the ordering is not uniform across the surface of the sphere, and there exists domains of high crystalline ordering separated by line defects with low or no ordering. 

The rest of the paper is organized as follows. In Section \ref{section_model}, we describe the model, the interaction potential, and the constraints used and the simulation details. In Section \ref{section_results}, we highlight the the main results --- the properties of the different phases, the nature of phase transitions between them and its dependence on shape anisotropy and the topological defects in the solid phase. In Section \ref{section_conclusion}, we discuss the results and their interpretations and implications, and conclude with possible future directions to this work.

\section{Model and Simulation Details} \label{section_model}

\begin{figure}
    \centering
    \includegraphics[width=\linewidth]{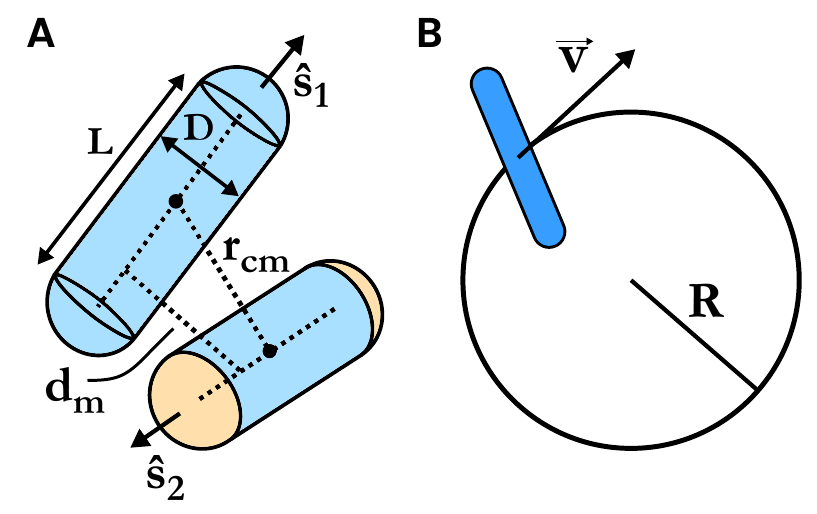}
    \caption{(A) Schematic diagram of the interaction between two SRS particles. The dashed line segment between the two end caps on each SRS is called the core. $\vu{s}_1,\ \vu{s}_2$ are the orientations of the two SRS respectively. $r_{\mathrm{cm}}$ is the distance between the center of mass of the two SRS while $d_m$ is the shortest distance between the cores. The force between these two particles depends on $d_m$ and acts along the shortest line segment between the two cores. The schematics apply to all pairs of particles. (B) Schematic of the spherical shell constraint on the SRS. The center of mass lies on the surface of the sphere, while its translational velocities is tangential to the surface.}
\label{fig:model}
\end{figure}

The system we study is a collection of soft repulsive spherocylinders (right circular cylinders with hemispherical end caps), each having mass $m$. The length of cylinder is $L$, and the diameter of the sphere as well as the cylinder is $D$. The shape anisotropy of such a molecule is $A=L/D$. These SRS particles interact with each other with a generalization of the Weeks-Chandler-Anderson potential \cite{weeks1971role} to non-spherical particles, in which the force acts along the line of shortest distance \cite{vega1994fast} between the cores of the spherocylinders, as opposed to between the line joining their centers. This interaction potential is given as follows:

\begin{align}
    U = \begin{cases}
            4\varepsilon\left[ \left( \frac{D}{d_m} \right)^{12} - \left( \frac{D}{d_m} \right)^6 \right] +\varepsilon, &d_m < 2^{1/6} D \\
            0 ,&d_m \geq 2^{1/6} D ,
        \end{cases}
\end{align}

where $d_m$ is the shortest distance between their axes (or cores), as shown in Fig. \ref{fig:model}A. The centers of mass of the spherocylinders (situated at their geometric centers) are constrained to lie on the surface of a sphere of radius $R$. The center of mass velocities are tangent to the surface of the sphere, whereas their orientation and angular velocities are unconstrained, as shown in Fig. \ref{fig:model}B. More specifically, the constraints are:

\begin{align} 
    \left|  \va{r}_i \right| &= R, \label{eqn:rconstr} 
    \\
    \va{v}_i \vdot \va{r}_i &= 0, 
\end{align}

where, $\va{r}_i,\ \va{v}_i$ are the center of mass position and velocity of the $i$th spherocylinder and the origin of coordinate system is at the center of the sphere. The constraints are applied to each $i$th spherocylinder.

\begin{figure*}
    \centering
    \includegraphics[width=\linewidth]{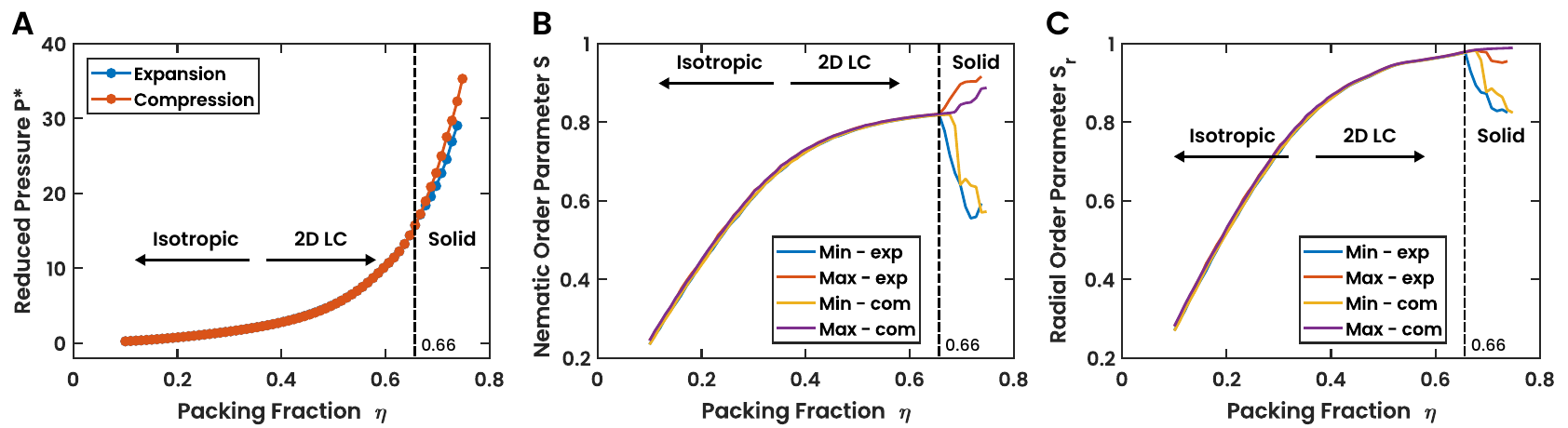}
    \caption{(A) Equation of state of the system, reduced pressure $P^*$ vs packing fraction $\eta$ of a system of 2500 particles with $A=5$, with both compression and expansion simulation schemes. (B) Nematic order parameters $S$ and (C) radial order parameter $S_r$ as a function $\eta$ for the same system and simulation schemes. The minimum and maximum are calculated over the 20 regions of the spherical surface. The closeness of the minimum and maximum indicates the degree of homogenous ordering across the system. A large difference between the maximum and minimum of the order parameters indicates inhomogenous ordering. The high (maximum) values of $S$ at high $\eta$ indicates the the appearance of crystalline domains, whereas the minimum values appear due to the defect lines that separate two domains . Therefore, the disagreement of the line of maximum and minimum nematic order parameter also is an indicator of appearance of solid phase. Both plots are with $T^*=5, A=5, N=2500$.}
    \label{fig:roundtrip}
\end{figure*}

We performed molecular dynamics (MD) simulations of this system in the constant number-volume-temperature (NVT) ensemble. We use velocity Verlet integration algorithm \cite{swope1982computer} to update the positions and velocities and an adaptation of the RATTLE algorithm \cite{andersen1983rattle} to enforce the constraints. All quantities, thermodynamic and structural are scaled by the system parameters $\varepsilon,\ D$ and calculated in reduced units: temperature $T^* = k_BT/\varepsilon$, pressure $P^* = aP/(k_BT)$, packing fraction $\eta = a\rho$, where $\rho = N/V$ is the density, $N$ is the number of particles, $V = 4\pi R^2$ is the surface area and $a = \pi D^2/4$ is the cross-sectional area of the spherocylinder. In our calculations, we take $k_B = 1$ and measure time in units of $D(m/\varepsilon)^{1/2}$. The temperature of the system was maintained using a Berendsen thermostat \cite{berendsen1984bath} with a temperature coupling time of $\tau_T = 0.05$ for smaller densities and down to $\tau_T = 0.01$ for larger densities. 

Because of the constraint equation \ref{eqn:rconstr}, the translational degrees of freedom for the particles is 2. Therefore, pressure is calculated as:
\begin{equation}
    P = \frac{1}{V}\left( NT + \frac{1}{2} \Xi \right)
\end{equation}
where $\Xi= \sum_{i=1}^N \va{r}_i\vdot\va{F}_i $ is the virial and $\va{F}_i$ is the force acting on the $i$th particle due to interaction with all other particles.
We prepared the initial state of the system with all particles evenly distributed on the surface of the sphere and having coordination number of 6, with the use of a Fibonacci sphere construction \cite{swinbank2006fibonacci}. Initially, all particle orientations $(\vu{s})$ are along the outward normal to the surface. The translational and rotational velocities are given random values in accordance with the constraints. We performed the simulations for a system 
size of $N = 2500$ particles and shape anisotropy $A = 5$. After setting up the initial state, we run the simulation at $T^* = 5$ for $4\times10^5$ timesteps to equilibrate the system. Following this, we simulated the system for another $4\times10^5$ timesteps while calculating and averaging the thermodynamic and structural quantities. We use an integration timestep of $\delta t = 0.001$ throughout the simulations. We simulate the system for a range of packing fractions ($\eta$) from 0.95 to 0.1. We changed the packing fraction after each stage of the simulation process (equilibriation and measurement) by changing the radius of the constraining sphere by an appropriate amount. To check for finite size effects, if any, we have also performed simulations for a system size of $N = 25000$. 

The ordering transitions are determined by calculating the nematic order parameter and the radial order parameter. The tensor order parameter is a traceless symmetric tensor $Q$ defined as:
\begin{equation} \label{eqn:nematic-order-parameter}
    Q_{\alpha \beta} = \frac{1}{N}\sum\limits_{i=1}^{N} \frac{3}{2} s_{i \alpha} s_{i \beta} \ -\ \frac{1}{2}\delta_{\alpha \beta}
\end{equation}
 where, $i,j$ corresponds to particle index and $\alpha,\beta$ corresponds to components of unit orientation vector $\vu{s}$. 
The scalar nematic order parameter $S$ is the largest eigenvalue of $Q$, and its corresponding (three-dimensional) eigenvector $\va{n}$ is the director of the ordered phase. In highly ordered states, $S\approx 1$ and in highly disordered states, $S\approx 0$. The radial order parameter quantifies how well the particles are aligned along their local radial direction and is defined as follows:
\begin{equation} \label{eqn:radial_order_parameter}
    S_r = \frac{3}{2} \frac{1}{N}\sum_{i=1}^N\vu{s}_i \vdot \vu{r}_i - \frac{1}{2}
\end{equation}

Since the system in consideration has spherical geometry, the nematic as well as radial order parameter can vary as a function of the position on the sphere. Therefore, we divide the system into 20 equally sized and shaped regions and calculate the order parameter for each of them separately. These 20 regions are the faces of an inscribed spherical icosahedron. 


The orientational ordering is also quantified with the orientational correlation of the particles, which is a function of the geodesic angle $\theta$, i.e. the angle subtended by the lines joining the center of the constraining sphere to two points on its surface. It is calculated as:
\begin{equation}
        OC(\theta) = \left< \qty[\frac{3}{2}(\vu{s}_i \vdot \vu{s}_j)^2 - \frac{1}{2}]\ \delta(\theta - \arccos{(\vu{r}_i \vdot \vu{r}_j)}) \right>,
\end{equation}
 where, the angle brackets indicate and average over all pairs of particles. The spatial ordering is quantified with the radial distribution function calculated as a function of the geodesic angle $\theta$ as follows:
\begin{equation}
    g(\theta) = \frac{V}{N^2}\sum_{i\ne j} \delta(\theta - \arccos{(\vu{r}_i \vdot \vu{r}_j)})
\end{equation}

The positional ordering is also quantified with the use of the structure factor $S(q)$ defined as follows:
\begin{equation}
    S(\va{q}) = \frac{1}{N}\sum_{i=1}^N\sum_{j=1}^{N} e^{-i\va{q}\vdot\left(\va{r}_i-\va{r}_j\right)}
\end{equation}

\section{Results} \label{section_results}
Below, we individually discuss the properties of each of the different phases observed in the system. The results and values  reported here are for the simulation of a system with $A=5.0,\ T^*=5.0$ and $N=2500$, unless otherwise stated.

\begin{figure}[t]
    \centering
    \includegraphics[width=8.3cm]{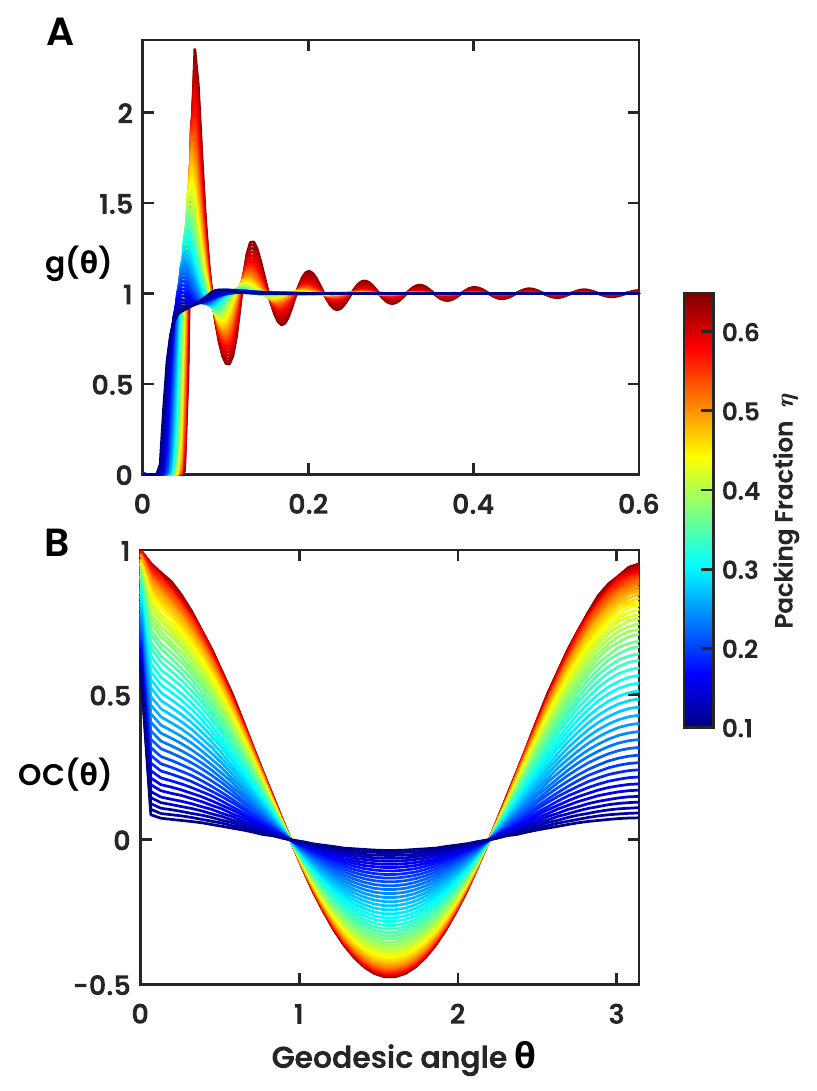}
    \caption{(A) Radial distribution function $g(\theta)$ as a function of the geodesic angle $\theta$ for low to medium packing fractions. At low packing fractions ($\eta \lesssim 0.35$), there is no structuring at all. But at medium packing fractions ($\eta \sim 0.35{-}0.65$), it shows short ranged spatial ordering and structure. (B) The orientational correlation $OC(\theta)$ as a function of the geodesic angle $\theta$ for same range of packing fractions. At low packing fractions, the orientations are not correlated and randomly oriented. At medium packing fractions, the sinusoidal orientational correlation shows that the particles are radially oriented.}
    \label{fig:radialdistribution-orientationalcorrelation}
\end{figure}

\begin{figure}[h]
    \centering
    \includegraphics[width=8.3cm]{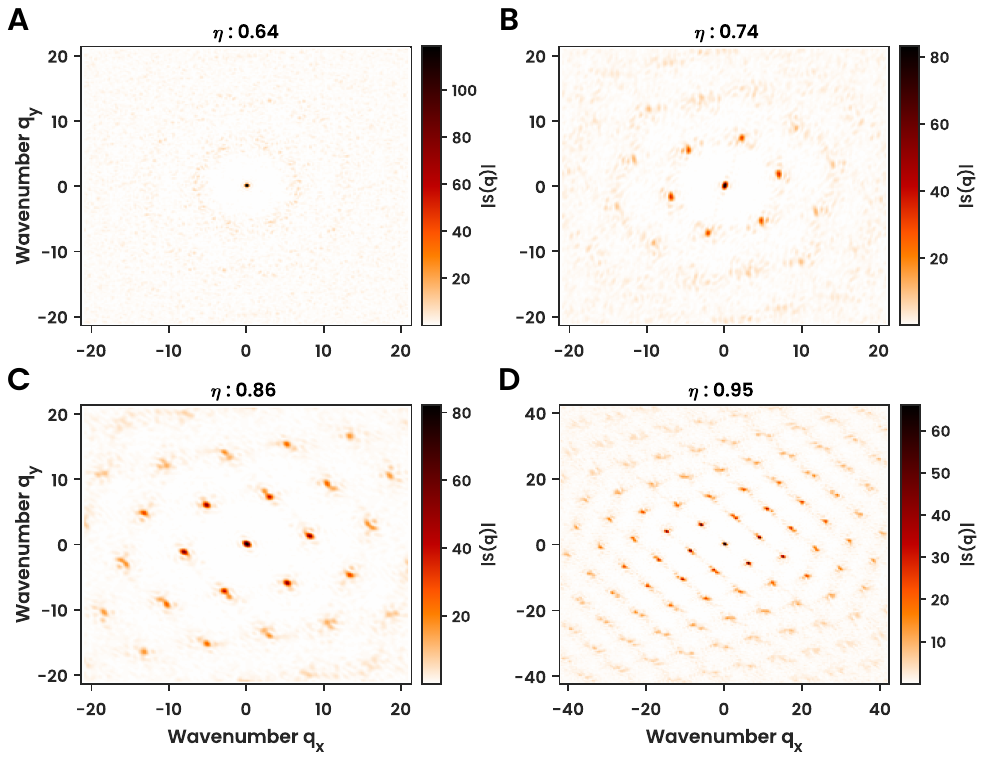}
    \caption{Emergence of positional ordering in the solid phase. (A) is the structure factor of a region in the nematic phase just below the phase transition point and it shows no long range ordering. (B), (C), (D) are the structure factors of a representative domain in the solid phase. As the packing fraction increases into the solid phase $\left(\eta \gtrsim 0.65\right)$, the range of the ordering also increases, indicated by the number of bright points in the plot. Before calculation of these structure factor, the region was first transformed such that its center of mass coincides with the origin and then flattened onto the $xy$ plane.}
    \label{fig:structurefactor}
\end{figure}

\begin{figure*}[t]
    \centering
    \includegraphics[width=15cm]{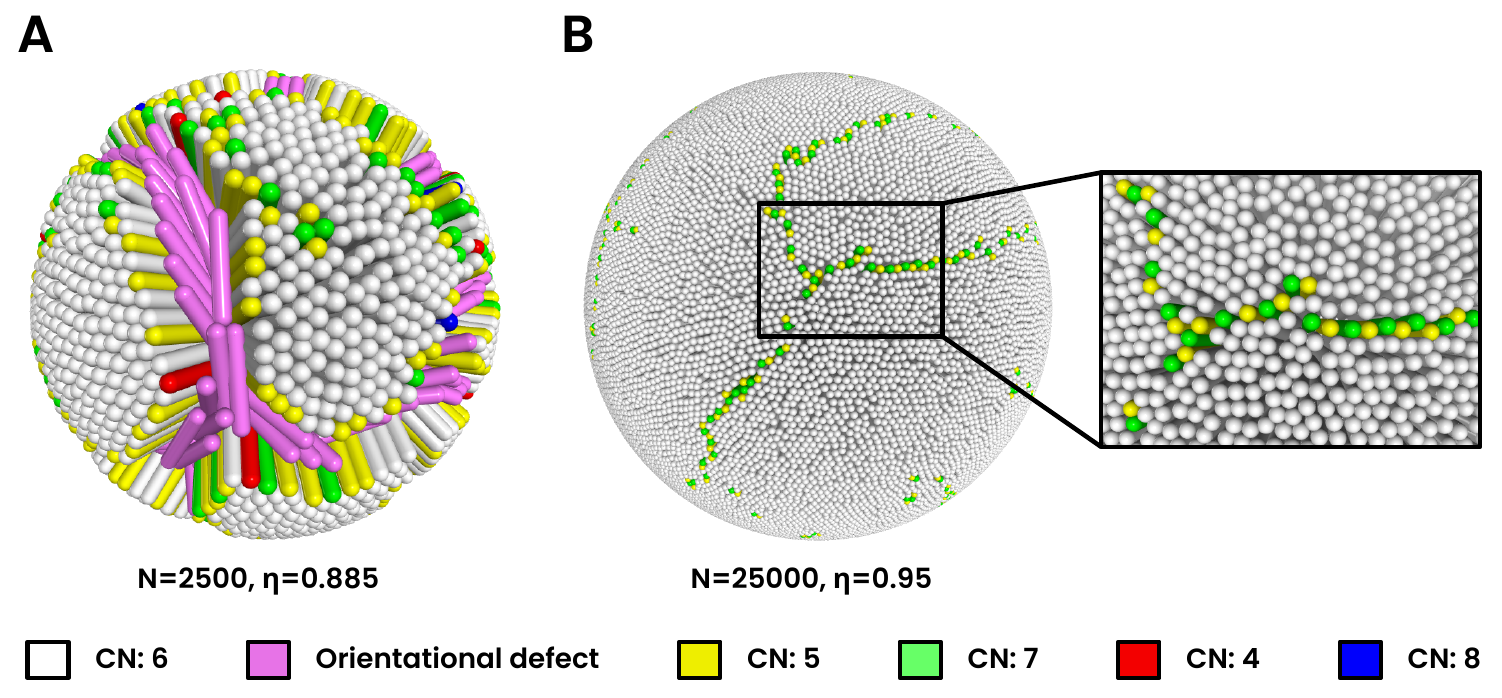}
    \caption{Defects seen in the solid phase of the system at two different system sizes (A) N=2500 particles and (B) N=25000 particles, with $A=5$. The particles are colored based on the coordination number (CN) and defect type. Both systems show large regions with the coordination number of 6, separated by defect lines, albeit of different types. The defect lines in the smaller system are formed of orientational defects as well as disclinations, while that of the larger system is formed by dislocation defects. They also form grain boundary scars.}
    \label{fig:defects}
\end{figure*}

\subsection{Fluid Phases}
At very low packing fractions $\left(\eta \lesssim 0.35\right)$, the system exhibits an isotropic fluid phase in which there is neither positional nor orientational order (Fig. \ref{fig:phases}A). The lack of positional order can be inferred from the radial distribution function $g(\theta)$ in Fig. \ref{fig:radialdistribution-orientationalcorrelation}A, which rises rapidly to 1 and saturates. The structure factor (Fig. \ref{fig:structurefactor}A) for $\eta<0.641$ also does not show any peaks other than the central $q=0$ peak, confirming the absence of positional ordering (see Fig. S1 in the supplementary information for a structure factor for $\eta=0.3$). The lack of orientational ordering can be inferred from the orientational correlation in Fig. \ref{fig:radialdistribution-orientationalcorrelation}B, nematic order parameter in Fig. \ref{fig:roundtrip}B, and radial order parameter in Fig. \ref{fig:roundtrip}C. The orientational correlation sharply decays to near 0 as $\theta$ increases from 0, and the nematic and radial order parameters take low values. We observe that the orientational ordering increases with the packing fraction.

At medium packing fractions ($\eta \sim 0.35{-}0.65 $), there is an emergence of substantial orientational ordering. This can be inferred from the  nematic and radial order parameters taking values ${\sim}0.8$ and ${\sim}1$, respectively (Fig. \ref{fig:roundtrip}B,C). The orientational ordering can also be understood from the form of the orientational correlation which tends to a sinusoidal-like curve as the packing fraction increases. Such a curve indicates that, on average, the particles are aligned along the local radial direction. The structure factor (Fig. \ref{fig:structurefactor}A) clearly shows that there is no positional ordering established at these packing fractions ($\eta \sim 0.35{-}0.65 $). The maximum and the minimum values of the nematic order parameter over the regions match closely, indicating a homogeneity in the orientational ordering across the spherical monolayer. Therefore, we term this phase as a radially oriented two-dimensional liquid crystal (2D LC). 

As the system transitions from isotropic fluid to 2D liquid crystal, all of its properties --- reduced pressure, nematic order parameter and radial order parameter --- change continuously and smoothly with packing fraction. This is in contrast to that seen in bulk 3D with SRS and hard spherocylinders (HSC) \cite{bolhuis_frenkel,cuetos2002monte,cuetos2005parsons}, where the equation of state changes discontinuously at the phase transition points. To determine whether the smooth transition from isotropic fluid to 2D liquid crystal is simply an artifact of the finite size of the system, we have performed simulations of a system of size $N=25000$ ie. 10 times the size of the initial system which we were studying. In this system as well, the transition from disordered to orientationally ordered LC phase is continuous (see Fig. S2 in the supplementary text for a plot of equation of state and order parameters for this system). Interestingly, this system does not show the divergence of maximum and minimum values of the nematic or radial order parameter within the calculated range of densities.

To determine whether the continuity in the isotropic to LC phase transition is seen only due to the fact that the simulation time is finite, we have also performed the simulations for a system of size $N=2500$ with both expanding and contracting schemes. That is, the system is simulated for each state point and then expanded to cover the range of packing fractions $\left(\eta: 0.95-0.1\right)$ Following the simulation of the lowest packing fraction $\left(\eta =0.1\right)$ state point, the system is then compressed to cover the same range in the reverse direction. The equation of state and order parameter for this simulation scheme are shown in Fig. \ref{fig:roundtrip}. If the transition is discontinuous, we expect to see a difference in the expansion and compression curves, like a hysteresis curve. However, in both these plots, we see that for the low $\left(\eta \lesssim 0.35\right)$ to medium packing fractions ($\eta \sim  0.35{-}0.65 $) range, the curve obtained by compression exactly follows the curve obtained by expansion
. The disagreement that is seen at higher packing fractions is due to the liquid crystal--solid phase transition.

Each spherocylinder in a 3D bulk system possesses up-down symmetry. Thus, when considering vectors along the rods' long axis, the order parameter cannot be associated with those vectors alone, as the negative of the vectors will also contribute equally to the order parameter. As a result, the order parameter is obtained from a symmetric traceless tensor that is compatible with the system's symmetries \cite{chaikin1995principles}. Though the particles in the constrained spherical system also exhibit the up-down symmetry, due to the presence of the constraining sphere, the environment inside and outside the sphere is different, i.e. the local density is higher inside the sphere than outside. This effect breaks the inversion symmetry at a macroscopic level. Therefore, the order parameter is vectorial in nature in this constrained system. The vectorial nature of the order parameter forbids the appearance of any cubic term in the Landau free energy expansion, as the free energy is a scalar quantity. As a result, the mean field isotropic to 2D LC phase transition is a continuous one. Continuous  isotropic--nematic transition was earlier reported in the amyloid fibril suspension under thermophoresis \cite{vigolo2017continuous}.

\subsection{The Solid Phase} \label{solid_phase_text}
The solid phase occurs at high packing fractions $(\eta\gtrsim 0.65)$, and shows high positional and orientational ordering within crystalline domains (Fig. \ref{fig:phases}C). These domains are separated by line defects having low or no ordering. To determine the nature of packing in the domains of the solid phase, we performed structure factor calculations on a representative domain of the state of the system for a range of packing fractions. Fig. \ref{fig:structurefactor} shows the emergence of well set translational ordering as the packing fractions is increased, as understood by the increasing number of rings in the plot. The structure factor (Fig. \ref{fig:structurefactor}) shows that the domains are crystalline in nature with hexagonal packing of the particles.  

The phase transition from 2D LC to solid phase is a first order transition, as indicated by the disagreement or hysteresis in the equation of state during the expansion and compression simulation regimes (Fig. \ref{fig:roundtrip}A). For sufficiently high shape anisotropy, the phase transition from 2D LC to solid is also identified by the divergence of maximum and minimum values of the order parameters across the spherical monolayer (Fig. \ref{fig:roundtrip}B,C). The LC to solid phase transition point depends on the anisotropy $A$ and temperature $T^*$. The phase transition also depends on the number of particles $N$ due to the finite-sized nature of the system. For an anisotropy of $A=5$, temperature of $T^*=5$, and system size of $N=2500$, the transition point occurs at $\eta \sim 0.66,\ P^* \sim 16.094$. 


\subsection{Topological Defects in the Solid Phase}
Defects in solids are known to exist in grain boundaries or at the edges of grains. These kind of defects exist only in polycrystalline solids and is generally absent in monocrystalline solids at low temperatures. However, in a spherical shell it is impossible to have a defect-free crystal due to its curvature. Moreover, it is due to these line defects that crystalline domains of high positional and orientational order can exist in the spherical shell i.e. some particles move out of the way to form line defects which allows the other particles to come closer and form tighter packing in a domain. At small system sizes and high spherical curvatures, the line defects consist of particles that are oriented at a large angle from the directors of the neighboring crystal domains and are nearly parallel to the surface of the constraining sphere (Fig. \ref{fig:defects}A). Therefore, the divergence of maximum and minimum of nematic order parameter of the regions implies the existence of a solid phase; however, the converse need not be true. In addition to these orientational line defects, there also exist disclinations (defects along the edges of the domains which have a coordination number other than six) (Fig. \ref{fig:defects}A). In addition to these defects at the boundaries of domains, there also exist point defects in the interior of domains. These are dislocations due to which one particle has a higher coordination number and another particle has a lower coordination number compared to their neighbors (Fig. \ref{fig:defects}A). However, at very large system sizes, the line defects are of a different nature. The orientational defects no longer occur. The defect lines separating domains consist only of positional dislocation defects (Fig. \ref{fig:defects}B). This occurs because the low curvature at high system sizes causes less frustration in the particles ordering and it can better accommodate crystalline packing. We also observe the appearance of grain boundary scars that appear and terminate within the spherical surface itself (Fig. \ref{fig:defects}B). In the limit of infinite system size, the curvature becomes zero and the system becomes an unconstrained 2D system, in which there can be perfect crystallinity.  

\begin{figure}[t]
    \centering
    \includegraphics[width=7cm]{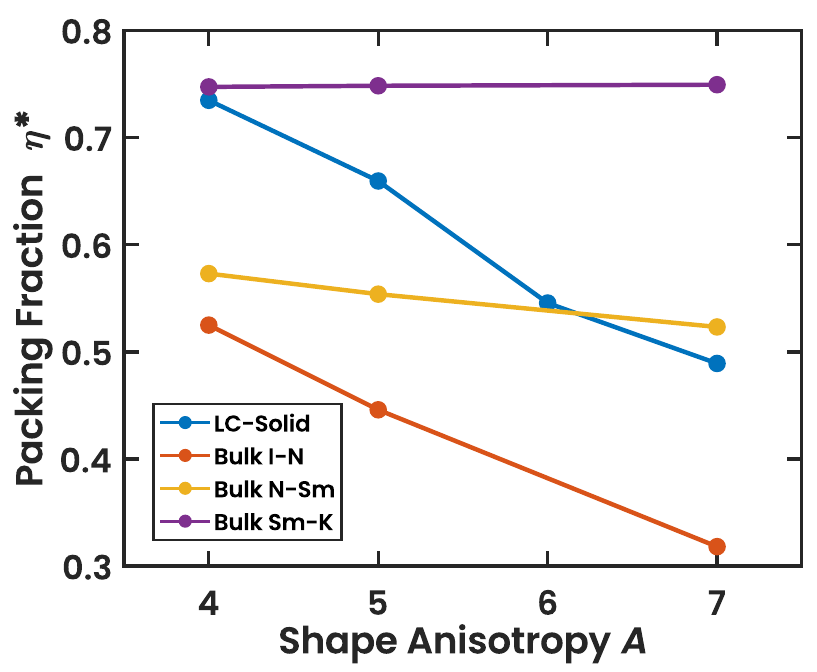}
    \caption{The shape anisotropy dependence of the phase boundary packing fraction $\eta*$ for the transition from the 2D liquid crystal to solid in for a spherical monolayer as well as the bulk transitions isotropic--nematic (I--N), nematic--smectic (N--Sm) and smectic--crystal (Sm--K). Packing fraction for SRS in bulk 3D is defined as $\eta=v_{hsc}\rho,\ \rho=N/V,\ v_{hsc}=\pi D^2(D/6+L/4)$. The data for the bulk transitions is taken from Cuetos and Martinez-Haya \cite{cuetos_haya}.}
    \label{fig:anisotropies}
\end{figure}

\begin{figure}[t]
    \centering
    \includegraphics[width=7cm]{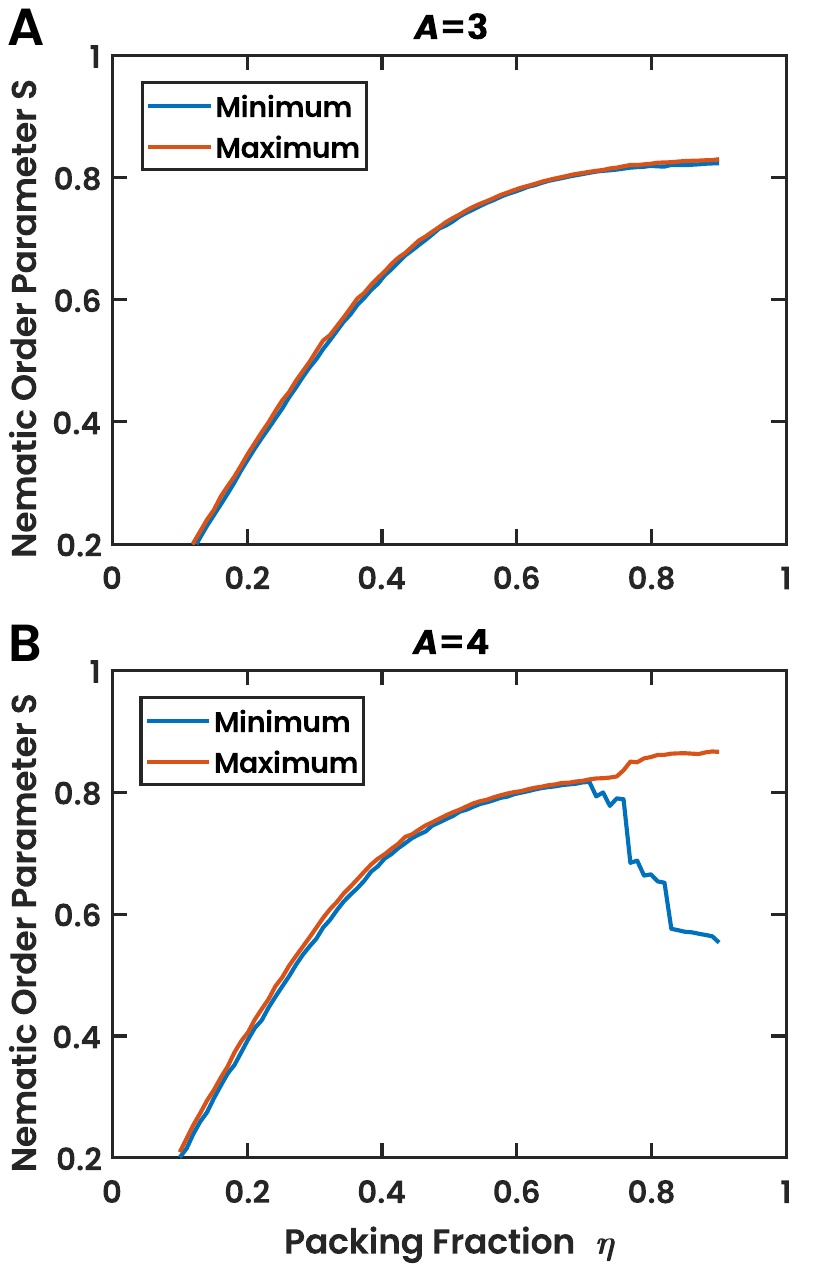}
    \caption{Nematic order parameter $S$ as a function of the packing fraction $\eta$ for (A) shape anisotropy,\(A=3\), and (B)\(A= 4\). Both are calculated for systems with $T^*=5,\ N=2500$. For $A=3$, there is no large difference of maximum and minimum values of the order parameters during the liquid crystal to solid phase transition while there is a large difference in the case of $A=4$. This indicates that there is indeed a critical shape anisotropy  $3 < A_c < 4$ below which the LC--solid transition is continuous.}
    \label{fig:LD34orderparameter}
\end{figure}

\subsection{Other Shape Anisotropies}
The phases of a system of a soft monolayer of spherocylinders naturally depend on the shape anisotropy of the rods in addition to the temperature, density and curvature of the constraining sphere. So far, we have looked at systems at a constant temperature of $T^* = 5 $, and constant shape anisotropy of $A = L/D = 5$ at varying packing fractions. To understand the dependence of the phases on the shape anisotropy, we have also simulated the system over the same range of packing fractions for various other shape anisotropies ($A$: 3--7). 

We observe that the phase transition from 2D LC to solid occurs at lower packing fractions as the shape anisotropy is increased, as shown by the decreasing curve in Fig. \ref{fig:anisotropies} (LC--solid). For example, for A = 4, the transition occurs for a packing fraction of ${\sim} 0.71$. In contrast, for A = 7, the transition happens at a packing fraction of ${\sim} 0.49$. This is expected because when the shape anisotropy increases, the tail end of the rods in the interior of the constraining sphere interact at closer distances and experience stronger repulsive forces. Due to this, the orientational defect lines would form at a lower packing fraction. This is the effect of the topological constraint. For bulk 3D cases, the critical density for the smectic to crystalline phase transition does not show similar dependence on the shape anisotropy of the particles. (Fig. \ref{fig:anisotropies} Bulk Sm--K). In 3D bulk, the smectic to crystalline transition is mainly controlled by the packing fraction or density. Interestingly, the 2D LC--solid phase transition packing fraction shows a similar decreasing behavior as the 3D bulk isotropic to nematic transition density.


Below a certain critical shape anisotropy $A_c$, orientational defects are no longer observed. This can be seen in Fig. \ref{fig:LD34orderparameter},(also see Fig. S3 in supplementary text) which shows that for $A=3$, the maximum and minimum nematic and radial order parameter lines match each other throughout, while for $A=4$, they deviate at high packing fractions. 
The homogeneity of nematic order parameter for $A=3$ indicates that there are no particles with large orientational deviations from its neighbors. This implies the absence of orientational defects. For $T^* = 5$ and $N=2500$ this critical shape anisotropy lies in the interval $A_c \in (3.47, 3.487)$. Below the critical shape anisotropy, the rods do not experience sufficient repulsive interactions on their tail ends in the interior of the sphere. Due to this, all of the rods can be accommodated in a radially aligned configuration and there is no need for the orientational defects to arise. However, below the critical shape anisotropy $A_c$, there is a gradual emergence of short-ranged (one or two rings in the structure factor) hexagonal positional order with increasing packing fractions (Fig. S4 in supplementary text). 
But, there is no large difference in the maximum and minimum values of nematic order parameter such as that seen above the critical shape anisotropy. 
We note that for shape anisotropy values lower than the critical value $(A<A_c)$, orientational defects, and hence a heterogeneity in the order may appear only if the packing fraction is extremely high ($ > 1$). However, such high packing fractions also result in extremely high pressures due to the sharply increasing pressure as a function of packing fraction (Fig. \ref{fig:roundtrip}A). Therefore, we limit our study to reasonable packing fractions in the range of 0--1. Moreover, in the limit that $A=L/D\rightarrow0$, the spherocylinder reduces to a simple sphere. In this limit, orientational defects are not possible even at arbitrarily large packing fractions. In 3D bulk, the presence of the nematic phase depends on the shape anisotropy and may not be observed below a minimum shape anisotropy \cite{cuetos_haya}. This is also seen in a system consisting of a mixture of active and passive spherocylinders, in which the presence of the nematic phase depends on the shape anisotropy as well as the relative activity \cite{jc2}. However, we observe the 2D LC phase for all values of $A$ that we considered in our simulations. i.e $A$: 3--7.

\section{Conclusions} \label{section_conclusion}
In this work, we have studied the phase behavior of soft spherocylinders whose center of masses are constrained to move on the surface of a sphere. We showed that the disordered fluid and orientationally ordered liquid crystalline phases appear at low and medium packing fractions respectively. The LC phase shows a hedgehog-particle like structure where the rods are aligned normal to the surface of the sphere. 
At even higher packing fractions, there appears a solid phase consisting of crystalline domains of high orientational and hexagonal close packing, separated by defect lines. We show that the transition from 2D liquid crystal to solid is a first order phase transition. This phase transition from liquid crystal to solid depends not only on the temperature and density of the system, but also on the curvature of the constraining spherical surface and the shape anisotropy of the rods. Keeping the other variables at fixed value, there appears a critical value of the shape anisotropy below which the 
orientational defects can no longer be observed. We also found that the LC--solid phase transition density (packing fraction) decreases with shape anisotropy of the particles, similar to the case of isotropic--nematic transition in 3D bulk. We observed various point and line defects in solid phase, which appear due to the curved geometry of the sphere. Interestingly, the point defects can have coordination number not only 5 and 7 (which is the case for spherical particles on spherical surface \cite{vest}) but also 4 and 8. For small system size with high curvature of the constraining sphere, there is appearance of orientational defects as well as disclinations, whereas for large system size with low curvature, we see dislocations and grain boundary scars emerging in the system.

We see several future directions of this work involving the examination of the phase behaviors of similar but modified systems. Constraining rods on other manifolds such as ellipsoids or toroids could give rise to distinct phases and defect structures due to the different properties of the manifolds. Chiral particles in 3D bulk show a twist deformation and cholesteric phases \cite{kamien2001order,dierking2014chiral}, but such particles in 2D cannot show twist and hence cholesteric phases are absent. However, if they are constrained in a spherical shell like in this work, it may be able to twist and show cholesteric phases. Spherocylinders in bulk show a nematic phase for packing fractions in the range of $\sim 0.5{-}0.6$ \cite{cuetos2002monte,jc} and more ordered smectic and crystal phases at higher packing fractions. This system of a spherical monolayer of spherocylinders also shows liquid crystalline phase around the same range of packing fractions. Therefore, it might be possible for a constrained spherical shell of spherocylinders in the ordered phase to exist in a bulk of unconstrained spherocylinders. In such a system the bulk particles close to the sphere will try to align with the local spherical director and particles far away try to be parallel. The transition from one to another is formed by defects and should be studied in detail. A system of active rods on such constraining manifolds, in which the non-equilibrium behavior of a mixture of active and passive rods on spherical geometry may give rise to novel structures and phase separations. Our future plan involves the studying of such systems.


\section*{Conflicts of interest}
There are no conflicts to declare.

\section*{Acknowledgements}
DR thanks KVPY, DST India for scholarship. JM thanks MHRD, India for fellowship. PKM thanks DST, India for financial support and SERB, India (IPA/2020/000034) and DAE, India for funding and computational support.



\balance


\bibliography{rsc} 

@article{bolhuis_frenkel,
  title={Tracing the phase boundaries of hard spherocylinders},
  author={Peter Bolhuis and Daan Frenkel},
  journal={The Journal of chemical physics},
  doi={10.1063/1.473404},
  year={1997},
  publisher={American Institute of Physics}
}

@article{allen,
  title={Isotropic-nematic interface of soft spherocylinders},
  author={Al-Barwani, Muataz S and Allen, Michael P},
  journal={Physical Review E},
  volume={62},
  number={5},
  pages={6706},
  year={2000},
  publisher={APS}
}

@article{bates_frenkel,
  title={Phase behavior of two-dimensional hard rod fluids},
  author={Bates, Martin A and Frenkel, Daan},
  journal={The Journal of Chemical Physics},
  volume={112},
  number={22},
  pages={10034--10041},
  year={2000},
  publisher={American institute of physics}
}

@article{dijkstra1,
  title={On the stability and finite-size effects of a columnar phase in single-component systems of hard-rod-like particles},
  author={Dussi, Simone and Chiappini, Massimiliano and Dijkstra, Marjolein},
  journal={Molecular Physics},
  volume={116},
  number={21-22},
  pages={2792--2805},
  year={2018},
  publisher={Taylor \& Francis}
}

@article{dijkstra2,
  title={Phase diagram of binary colloidal rod-sphere mixtures from a 3D real-space analysis of sedimentation--diffusion equilibria},
  author={Bakker, Henri{\"e}tte E and Dussi, Simone and Droste, Barbera L and Besseling, Thijs H and Kennedy, Chris L and Wiegant, Evert I and Liu, Bing and Imhof, Arnout and Dijkstra, Marjolein and van Blaaderen, Alfons},
  journal={Soft Matter},
  volume={12},
  number={45},
  pages={9238--9245},
  year={2016},
  publisher={Royal Society of Chemistry}
}

@article{dijkstra3,
  title={Phase behavior of binary mixtures of thick and thin hard rods},
  author={van Roij, Ren{\'e} and Mulder, Bela and Dijkstra, Marjolein},
  journal={Physica A: Statistical Mechanics and its Applications},
  volume={261},
  number={3-4},
  pages={374--390},
  year={1998},
  publisher={Elsevier}
}

@article{adams_rod_sphere_expt,
  title={Entropically driven microphase transitions in mixtures of colloidal rods and spheres},
  author={Adams, Marie and Dogic, Zvonimir and Keller, Sarah L and Fraden, Seth},
  journal={Nature},
  volume={393},
  number={6683},
  pages={349--352},
  year={1998},
  publisher={Nature Publishing Group}
}

@article{glaser,
 title={Phase behavior of polarizable spherocylinders in external fields},
  author={Rotunno, Melissa and Bellini, Tommaso and Lansac, Yves and Glaser, Matthew A},
  journal={The Journal of chemical physics},
  volume={121},
  number={11},
  pages={5541--5549},
  year={2004},
  publisher={American Institute of Physics}
}

@article{jc,
  title = {Heating leads to liquid-crystal and crystalline order in a two-temperature active fluid of rods},
  author = {Chattopadhyay, Jayeeta and Pannir-Sivajothi, Sindhana and Varma, Kaarthik and Ramaswamy, Sriram and Dasgupta, Chandan and Maiti, Prabal K.},
  journal = {Phys. Rev. E},
  volume = {104},
  issue = {5},
  pages = {054610},
  numpages = {15},
  year = {2021},
  month = {Nov},
  publisher = {American Physical Society},
  doi = {10.1103/PhysRevE.104.054610},
  url = {https://link.aps.org/doi/10.1103/PhysRevE.104.054610}
}

@misc{jc2,
    title={Two-temperature activity induces liquid-crystal phases inaccessible in equilibrium},
    author={Jayeeta Chattopadhyay and Sriram Ramaswamy and Chandan Dasgupta and Prabal K. Maiti},
    year={2022},
    eprint={2205.00667},
    archivePrefix={arXiv},
    primaryClass={cond-mat.soft}
}

@article{lubensky,
  title={Orientational order and vesicle shape},
  author={T. Lubensky and Jacques Prost},
  journal={J Phys II},
  volume={2},
  pages={371},
  year={1992},
}

@article{Bates,
  title={Nematic ordering and defects on the surface of a sphere: A Monte Carlo simulation study},
  author={Bates, Martin A},
  journal={The Journal of chemical physics},
  volume={128},
  number={10},
  pages={104707},
  year={2008},
  publisher={American Institute of Physics}
}

@article{cuetos2002monte,
  title={Monte Carlo study of liquid crystal phases of hard and soft spherocylinders},
  author={Cuetos, Alejandro and Mart{\i}nez-Haya, B and Rull, LF and Lago, S},
  journal={The Journal of chemical physics},
  volume={117},
  number={6},
  pages={2934--2946},
  year={2002},
  publisher={American Institute of Physics}
}

@article{cuetos2005parsons,
  title={Parsons- Lee and Monte Carlo Study of Soft Repulsive Nematogens},
  author={Cuetos, Alejandro and Martinez-Haya, Bruno and Lago, S and Rull, LF},
  journal={The Journal of Physical Chemistry B},
  volume={109},
  number={28},
  pages={13729--13736},
  year={2005},
  publisher={ACS Publications}
}

@article{smallenburg,
  title={Close packing of rods on spherical surfaces},
  author={Smallenburg, Frank and L{\"o}wen, Hartmut},
  journal={The Journal of Chemical Physics},
  volume={144},
  number={16},
  pages={164903},
  year={2016},
  publisher={AIP Publishing LLC}
}

@article{exotic_structure,
  title={Smectic monolayer confined on a sphere: topology at the particle scale},
  author={Allahyarov, Elshad and Voigt, Axel and L{\"o}wen, Hartmut},
  journal={Soft Matter},
  volume={13},
  number={44},
  pages={8120--8135},
  year={2017},
  publisher={Royal Society of Chemistry}
}

@article{dhakal,
  title={Nematic liquid crystals on spherical surfaces: Control of defect configurations by temperature, density, and rod shape},
  author={Dhakal, Subas and Solis, Francisco J and De La Cruz, Monica Olvera},
  journal={Physical Review E},
  volume={86},
  number={1},
  pages={011709},
  year={2012},
  publisher={APS}
}

@article{shin,
  title={Topological defects in spherical nematics},
  author={Shin, Homin and Bowick, Mark J and Xing, Xiangjun},
  journal={Physical review letters},
  volume={101},
  number={3},
  pages={037802},
  year={2008},
  publisher={APS}
}

@article{defect_expt1,
  title={Nematic-smectic transition in spherical shells},
  author={Lopez-Leon, Teresa and Fernandez-Nieves, Alberto and Nobili, Maurizio and Blanc, Christophe},
  journal={Physical Review Letters},
  volume={106},
  number={24},
  pages={247802},
  year={2011},
  publisher={APS}
}

@article{cuetos_haya,
  title={Liquid crystal phase diagram of soft repulsive rods and its mapping on the hard repulsive reference fluid},
  author={Alejandro Cuetos and Bruno Martinez-Haya},
  journal={Molecular Physics},
  doi={10.1080/00268976.2014.996191},
  year={2015},
  publisher={ Taylor & Francis}
}

@article{janssen_kaiser_lowen,
 title={Aging and rejuvenation of active matter under topological constraints},
  author={Janssen, Liesbeth and Kaiser, Andreas and L{\"o}wen, Hartmut},
  journal={Scientific reports},
  volume={7},
  number={1},
  pages={1--13},
  year={2017},
  publisher={Nature Publishing Group}
}

@article{bio2,
  title={Clonal analysis of patterns of growth, stem cell activity, and cell movement during the development and maintenance of the murine corneal epithelium},
  author={Collinson, J Martin and Morris, Lucy and Reid, Alasdair I and Ramaesh, Thaya and Keighren, Margaret A and Flockhart, Jean H and Hill, Robert E and Tan, Seong-Seng and Ramaesh, Kanna and Dhillon, Baljean and others},
  journal={Developmental dynamics: an official publication of the American Association of Anatomists},
  volume={224},
  number={4},
  pages={432--440},
  year={2002},
  publisher={Wiley Online Library}
}

@article{bio1,
   title={Reconstruction of zebrafish early embryonic development by scanned light sheet microscopy},
  author={Keller, Philipp J and Schmidt, Annette D and Wittbrodt, Joachim and Stelzer, Ernst HK},
  journal={science},
  volume={322},
  number={5904},
  pages={1065--1069},
  year={2008},
  publisher={American Association for the Advancement of Science}
}

@article{colloidosome1,
  title={Colloidal spheres confined by liquid droplets: Geometry, physics, and physical chemistry},
  author={Manoharan, Vinothan N},
  journal={Solid state communications},
  volume={139},
  number={11-12},
  pages={557--561},
  year={2006},
  publisher={Elsevier}
}

@article{colloidosome2,
  title={Double emulsion-templated nanoparticle colloidosomes with selective permeability},
  author={Lee, Daeyeon and Weitz, David A},
  journal={Advanced Materials},
  volume={20},
  number={18},
  pages={3498--3503},
  year={2008},
  publisher={Wiley Online Library}
}

@article{microfluidics1,
  title={Liquid crystals in curved confined geometries: Microfluidics bring new capabilities for photonic applications and beyond},
  author={Chen, Han-Qing and Wang, Xi-Yuan and Bisoyi, Hari Krishna and Chen, Lu-Jian and Li, Quan},
  journal={Langmuir},
  volume={37},
  number={13},
  pages={3789--3807},
  year={2021},
  publisher={ACS Publications}
}

@article{thick_nematic_shell,
  title={Novel defect structures in nematic liquid crystal shells},
  author={Fern{\'a}ndez-Nieves, Alberto and Vitelli, Vincenzo and Utada, Andrew S and Link, Darren R and M{\'a}rquez, Manuel and Nelson, David R and Weitz, David A},
  journal={Physical review letters},
  volume={99},
  number={15},
  pages={157801},
  year={2007},
  publisher={APS}
}

@article{nematic_expt1,
  title={Thermally switched release from nanoparticle colloidosomes},
  author={Zhou, Shaobing and Fan, Jing and Datta, Sujit S and Guo, Ming and Guo, Xing and Weitz, David A},
  journal={Advanced Functional Materials},
  volume={23},
  number={47},
  pages={5925--5929},
  year={2013},
  publisher={Wiley Online Library}
}

@article{nematic_expt2,
  title={Frustrated nematic order in spherical geometries},
  author={Lopez-Leon, Teresa and Koning, V and Devaiah, KBS and Vitelli, Vincenzo and Fernandez-Nieves, A},
  journal={Nature Physics},
  volume={7},
  number={5},
  pages={391--394},
  year={2011},
  publisher={Nature Publishing Group}
}

@article{andersen1983rattle,
  title={Rattle: A “velocity” version of the shake algorithm for molecular dynamics calculations},
  author={Andersen, Hans C},
  journal={Journal of computational Physics},
  volume={52},
  number={1},
  pages={24--34},
  year={1983},
  publisher={Elsevier}
}

@article{swope1982computer,
  title={A computer simulation method for the calculation of equilibrium constants for the formation of physical clusters of molecules: Application to small water clusters},
  author={Swope, William C and Andersen, Hans C and Berens, Peter H and Wilson, Kent R},
  journal={The Journal of chemical physics},
  volume={76},
  number={1},
  pages={637--649},
  year={1982},
  publisher={American Institute of Physics}
}

@article{berendsen1984bath,
  title={Molecular dynamics with coupling to an external bath},
  author={Berendsen, Herman JC and Postma, JPM van and Van Gunsteren, Wilfred F and DiNola, ARHJ and Haak, Jan R},
  journal={The Journal of chemical physics},
  volume={81},
  number={8},
  pages={3684--3690},
  year={1984},
  publisher={American Institute of Physics}
}

@article{weeks1971role,
  title={Role of repulsive forces in determining the equilibrium structure of simple liquids},
  author={Weeks, John D and Chandler, David and Andersen, Hans C},
  journal={The Journal of chemical physics},
  volume={54},
  number={12},
  pages={5237--5247},
  year={1971},
  publisher={American Institute of Physics}
}

@article{vest,
  title={Dynamics of a monodisperse Lennard-Jones system on a sphere},
  author={Vest, Julien-Piera and Tarjus, Gilles and Viot, Pascal},
  journal={Molecular Physics},
  volume={112},
  number={9-10},
  pages={1330--1335},
  year={2014},
  publisher={Taylor \& Francis}
}

@article{swinbank2006fibonacci,
  title={Fibonacci grids: A novel approach to global modelling},
  author={Swinbank, Richard and James Purser, R},
  journal={Quarterly Journal of the Royal Meteorological Society: A journal of the atmospheric sciences, applied meteorology and physical oceanography},
  volume={132},
  number={619},
  pages={1769--1793},
  year={2006},
  publisher={Wiley Online Library}
}

@article {devries,
 title={Divalent metal nanoparticles},
  author={DeVries, Gretchen A and Brunnbauer, Markus and Hu, Ying and Jackson, Alicia M and Long, Brenda and Neltner, Brian T and Uzun, Oktay and Wunsch, Benjamin H and Stellacci, Francesco},
  journal={Science},
  volume={315},
  number={5810},
  pages={358--361},
  year={2007},
  publisher={American Association for the Advancement of Science}
}

@article{onsager1949effects,
  title={The effects of shape on the interaction of colloidal particles},
  author={Onsager, Lars},
  journal={Annals of the New York Academy of Sciences},
  volume={51},
  number={4},
  pages={627--659},
  year={1949},
  publisher={Blackwell Publishing Ltd Oxford, UK}
}

@book{de1993physics,
  title={The physics of liquid crystals},
  author={De Gennes, Pierre-Gilles and Prost, Jacques},
  number={83},
  year={1993},
  publisher={Oxford university press}
}

@article{poulin1997novel,
  title={Novel colloidal interactions in anisotropic fluids},
  author={Poulin, Philippe and Stark, Holger and Lubensky, TC and Weitz, DA},
  journal={Science},
  volume={275},
  number={5307},
  pages={1770--1773},
  year={1997},
  publisher={American Association for the Advancement of Science}
}

@article{liu2013nematic,
  title={Nematic liquid crystal boojums with handles on colloidal handlebodies},
  author={Liu, Qingkun and Senyuk, Bohdan and Tasinkevych, Mykola and Smalyukh, Ivan I},
  journal={Proceedings of the National Academy of Sciences},
  volume={110},
  number={23},
  pages={9231--9236},
  year={2013},
  publisher={National Acad Sciences}
}

@article{senyuk2013topological,
  title={Topological colloids},
  author={Senyuk, Bohdan and Liu, Qingkun and He, Sailing and Kamien, Randall D and Kusner, Robert B and Lubensky, Tom C and Smalyukh, Ivan I},
  journal={Nature},
  volume={493},
  number={7431},
  pages={200--205},
  year={2013},
  publisher={Nature Publishing Group}
}

@article{nych2013assembly,
  title={Assembly and control of 3D nematic dipolar colloidal crystals},
  author={Nych, Andriy and Ognysta, Ulyana and {\v{S}}karabot, Miha and Ravnik, Miha and {\v{Z}}umer, Slobodan and Mu{\v{s}}evi{\v{c}}, Igor},
  journal={Nature communications},
  volume={4},
  number={1},
  pages={1--8},
  year={2013},
  publisher={Nature Publishing Group}
}

@article{musevic2006two,
  title={Two-dimensional nematic colloidal crystals self-assembled by topological defects},
  author={Musevic, Igor and Skarabot, Miha and Tkalec, Uros and Ravnik, Miha and Zumer, Slobodan},
  journal={Science},
  volume={313},
  number={5789},
  pages={954--958},
  year={2006},
  publisher={American Association for the Advancement of Science}
}

@book{chaikin1995principles,
  title={Principles of condensed matter physics},
  author={Chaikin, Paul M and Lubensky, Tom C and Witten, Thomas A},
  volume={10},
  year={1995},
  publisher={Cambridge university press Cambridge}
}

@article{nelson_sp3,
  title={Toward a Tetravalent Chemistry of Colloids},
  author={Nelson, David R.},
  journal={Nano Letters},
  volume={2},
  number={10},
  pages={1125-1129},
  year={2002}
}

@article{hedgehog,
  title={Anomalous dispersions of ‘hedgehog’ particles},
  author={Joong Hwan Bahng and Bongjun Yeom and Yichun Wang and Siu On Tung and J. Damon Hoff and Nicholas Kotov},
  journal={Nature},
  volume={517},
  pages={596–599},
  year={2015}
}

@article{poincare1885courbes,
  title={Sur les courbes d{\'e}finies par les {\'e}quations diff{\'e}rentielles},
  author={Poincar{\'e}, Henri},
  journal={J. Math. Pures Appl.},
  volume={4},
  pages={167--244},
  year={1885}
}

@article{Brouwer1912,
author = {Brouwer, L.E.J.},
journal = {Mathematische Annalen},
language = {ger},
pages = {97-115},
title = {Über Abbildung von Mannigfaltigkeiten},
url = {http://eudml.org/doc/158520},
volume = {71},
year = {1912},
}

@article{vigolo2017continuous,
  title={Continuous isotropic-nematic transition in amyloid fibril suspensions driven by thermophoresis},
  author={Vigolo, Daniele and Zhao, Jianguo and Handschin, Stephan and Cao, Xiaobao and deMello, Andrew J and Mezzenga, Raffaele},
  journal={Scientific reports},
  volume={7},
  number={1},
  pages={1--7},
  year={2017},
  publisher={Nature Publishing Group}
}

@article{mcgrother1996re,
  title={A re-examination of the phase diagram of hard spherocylinders},
  author={McGrother, Simon C and Williamson, Dave C and Jackson, George},
  journal={The Journal of chemical physics},
  volume={104},
  number={17},
  pages={6755--6771},
  year={1996},
  publisher={American Institute of Physics}
}

@article{maiti2002induced,
  title={Induced anticlinic ordering and nanophase segregation of bow-shaped molecules in a smectic solvent},
  author={Maiti, Prabal K and Lansac, Yves and Glaser, Matthew A and Clark, Noel A},
  journal={Physical review letters},
  volume={88},
  number={6},
  pages={065504},
  year={2002},
  publisher={APS}
}

@article{lansac2003phase,
  title={Phase behavior of bent-core molecules},
  author={Lansac, Yves and Maiti, Prabal K and Clark, Noel A and Glaser, Matthew A},
  journal={Physical Review E},
  volume={67},
  number={1},
  pages={011703},
  year={2003},
  publisher={APS}
}

@article{vega1994fast,
  title={A fast algorithm to evaluate the shortest distance between rods},
  author={Vega, Carlos and Lago, Santiago},
  journal={Computers \& chemistry},
  volume={18},
  number={1},
  pages={55--59},
  year={1994},
  publisher={Elsevier}
}

@article{earl2001computer,
  title={Computer simulations of soft repulsive spherocylinders},
  author={Earl, David J and Ilnytskyi, Jaroslav and Wilson, Mark R},
  journal={Molecular physics},
  volume={99},
  number={20},
  pages={1719--1726},
  year={2001},
  publisher={Taylor \& Francis}
}

@article{dussi2016entropy,
  title={Entropy-driven formation of chiral nematic phases by computer simulations},
  author={Dussi, Simone and Dijkstra, Marjolein},
  journal={Nature communications},
  volume={7},
  number={1},
  pages={1--10},
  year={2016},
  publisher={Nature Publishing Group}
}

@article{dussi2018hard,
  title={Hard competition: Stabilizing the elusive biaxial nematic phase in suspensions of colloidal particles with extreme lengths},
  author={Dussi, Simone and Tasios, Nikos and Drwenski, Tara and Van Roij, Ren{\'e} and Dijkstra, Marjolein},
  journal={Physical Review Letters},
  volume={120},
  number={17},
  pages={177801},
  year={2018},
  publisher={APS}
}

@article{cuetos2003liquid,
  title={Liquid crystal behavior of the Kihara fluid},
  author={Cuetos, Alejandro and Martinez-Haya, Bruno and Lago, S and Rull, LF},
  journal={Physical Review E},
  volume={68},
  number={1},
  pages={011704},
  year={2003},
  publisher={APS}
}

@article{patti2012brownian,
  title={Brownian dynamics and dynamic Monte Carlo simulations of isotropic and liquid crystal phases of anisotropic colloidal particles: A comparative study},
  author={Patti, Alessandro and Cuetos, Alejandro},
  journal={Physical Review E},
  volume={86},
  number={1},
  pages={011403},
  year={2012},
  publisher={APS}
}

@article{martinez2007stability,
  title={Stability of nematic and smectic phases in rod-like mesogens with orientation- dependent attractive interactions},
  author={Mart{\'\i}nez-Haya, B and Cuetos, A},
  journal={The Journal of Physical Chemistry B},
  volume={111},
  number={28},
  pages={8150--8157},
  year={2007},
  publisher={ACS Publications}
}

@article{marechal2011phase,
  title={Phase behavior of hard colloidal platelets using free energy calculations},
  author={Marechal, Matthieu and Cuetos, Alejandro and Mart{\'\i}nez-Haya, Bruno and Dijkstra, Marjolein},
  journal={The Journal of chemical physics},
  volume={134},
  number={9},
  pages={094501},
  year={2011},
  publisher={American Institute of Physics}
}

@article{van1995transverse,
  title={Transverse interlayer order in lyotropic smectic liquid crystals},
  author={van Roij, Ren{\'e} and Bolhuis, Peter and Mulder, Bela and Frenkel, Daan},
  journal={Physical Review E},
  volume={52},
  number={2},
  pages={R1277},
  year={1995},
  publisher={APS}
}

@article{dierking2014chiral,
  title={Chiral liquid crystals: structures, phases, effects},
  author={Dierking, Ingo},
  journal={Symmetry},
  volume={6},
  number={2},
  pages={444--472},
  year={2014},
  publisher={Multidisciplinary Digital Publishing Institute}
}

@article{kamien2001order,
  title={Order and frustration in chiral liquid crystals},
  author={Kamien, Randall D and Selinger, Jonathan V},
  journal={Journal of Physics: Condensed Matter},
  volume={13},
  number={3},
  pages={R1},
  year={2001},
  publisher={IOP Publishing}
}

@article{law2020phase,
  title={Phase transitions on non-uniformly curved surfaces: coupling between phase and location},
  author={Law, Jack O and Dean, Jacob M and Miller, Mark A and Kusumaatmaja, Halim},
  journal={Soft Matter},
  volume={16},
  number={34},
  pages={8069--8077},
  year={2020},
  publisher={Royal Society of Chemistry}
}

@article{law2018nucleation,
  title={Nucleation on a sphere: the roles of curvature, confinement and ensemble},
  author={Law, Jack O and Wong, Alex G and Kusumaatmaja, Halim and Miller, Mark A},
  journal={Molecular Physics},
  volume={116},
  number={21-22},
  pages={3008--3019},
  year={2018},
  publisher={Taylor \& Francis}
}

@article{akram2020chiral,
  title={Chiral molecules on curved colloidal membranes},
  author={Akram, Sk Ashif and Behera, Arabinda and Sharma, Prerna and Sain, Anirban},
  journal={Soft Matter},
  volume={16},
  number={45},
  pages={10310--10319},
  year={2020},
  publisher={Royal Society of Chemistry}
}
\bibliographystyle{rsc} 

\end{document}


\maketitle

This supplementary material contains results to show that 1. there is no positional ordering in the isotropic phase, 2. the continuous transition from isotropic to 2D liquid crystal is not just a finite size effect, 3. shape anisotropy has a similar effect on the radial order parameter as compared to the nematic order parameter, and 4. there can be short ranged positional ordering below the critical anisotropy $A_c$.

{\large \textit{\textbf{1. No positional ordering in the isotropic phase}}}

\begin{figure}[h]
    \centering
    \includegraphics[width=0.5\linewidth]{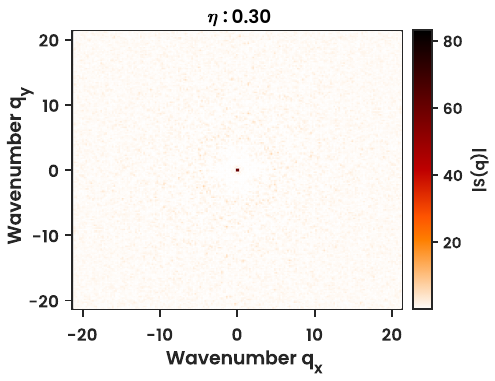}
    \caption{Structure factor for a packing fraction of $\eta=0.3$ for a system of $N=2500,\ T^*=5,\ A=5$. The absence of any peak apart from the $q=0$ peak confirms the lack of positional ordering at this and lower densities.}
    \label{fig:structurefactor_isotropic}
\end{figure}

As is mentioned in the main text, the system shows different phases depending on the density. These phases show different positional and orientational ordering. We have showed that in the 2D LC phase, there is orientational ordering but no positional ordering. Through the nematic and radial order parameters, and orientational correlation we have then shown that there is no orientational ordering in the isotropic phase as well. While the radial distribution function in the main text indicates that there is no positional ordering in the isotropic phase, we confirm this fact by the calculation of structure factor for the upper range of density for the isotropic phase (Fig. \ref{fig:structurefactor_isotropic}). The structure factor does not show any other peak apart from the one at $q=0$. This confirms our observation of lack of positional ordering in the isotropic phase. 

{\large \textit{\textbf{2.  No finite size effects on the isotropic--LC phase transition}}}

\begin{figure}[h]
    \centering
    \includegraphics[width=\linewidth]{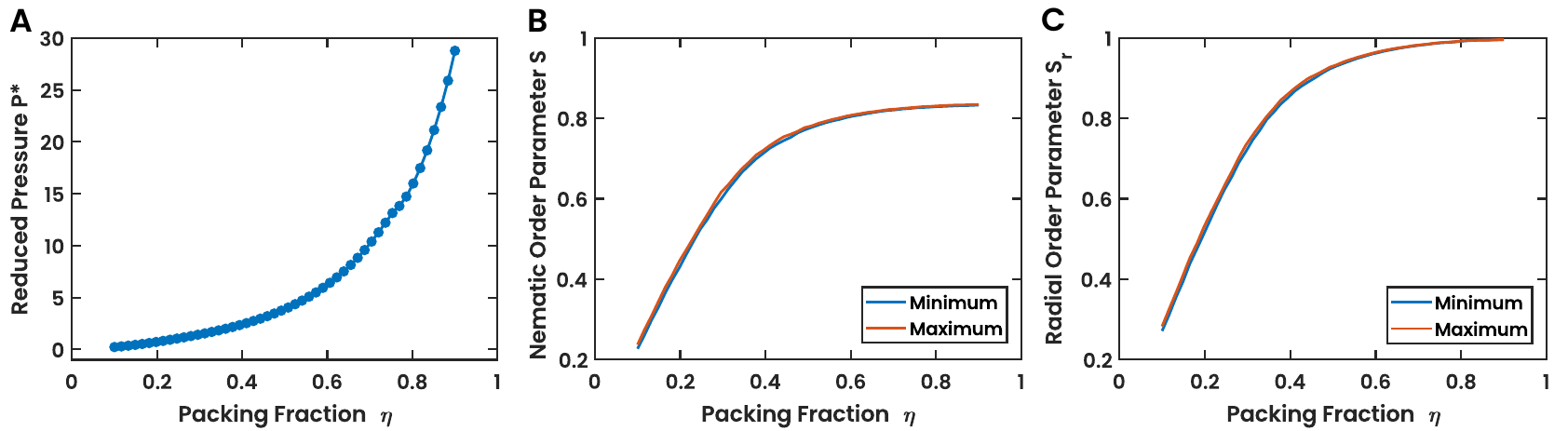}
    \caption{(A) Equation of state of the system i.e. reduced pressure $P^*$ vs packing fraction $\eta$ of a system of 25000 particles with $A=5,\ T^*=5$. (B) Nematic order parameter $S$ and (C) radial order parameter $S_r$ as a function of the packing fraction $\eta$ for the same system.}
    \label{fig:largesystem}
\end{figure}

To clarify whether the continuous transition from isotropic to 2D LC phase occurs because of the fact that our system size is finite, ($N=2500$ number of particles in the system), we simulated the system with a much larger number of particles i.e. $N=25000$. We observe that the isotropic to liquid crystalline phase transition is continuous with respect to the equation of state and order parameters in this case as well (Fig \ref{fig:largesystem}), confirming that the nature of such transition from isotropic to LC is a continuous one for a spherical monolayer of SRS particles, irrespective of the system size.  

{\large \textit{\textbf{3. Radial order parameter shows similar nature as nematic order parameter for different shape anisotropy}}}

\begin{figure}[h]
    \centering
    \includegraphics[width=0.75\linewidth]{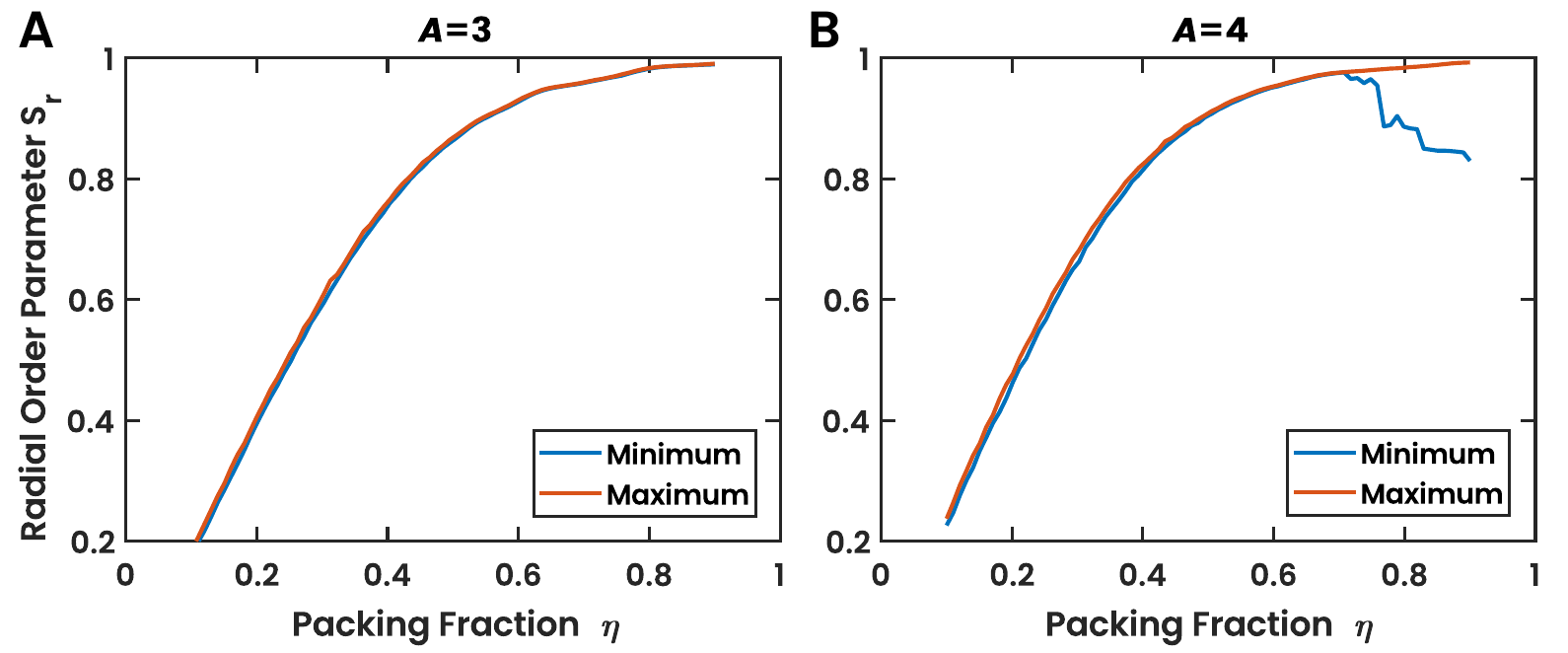}
    \caption{Radial order parameter $S_r$ as a function of packing fraction $\eta$ for shape anisotropies of (A) $A=3$, and (B) $A=4$. Both are calculated for systems with $T^*=5,\ N=2500$.}
    \label{fig:radialorderparameterA34}
\end{figure}

For shape anisotropy $A=3$, the maximum and minimum values of radial order parameter is same over the surface of the sphere, indicating homogeneous ordering of the spherocylinders in the full range of densities (Fig. \ref{fig:radialorderparameterA34}A). Whereas, for $A=4$, orientational defects arise at high densities and there is a large difference between the maximum and minimum values of radial order parameter $S_r$ over the surface of the sphere (Fig. \ref{fig:radialorderparameterA34}B). This shows that there exists a critical shape anisotropy $A_c$ below which orientational defects can no longer arise.

\pagebreak

{\large \textit{\textbf{4. Positional ordering below $A_c$}}}

\begin{figure}
    \centering
    \includegraphics[width=\linewidth]{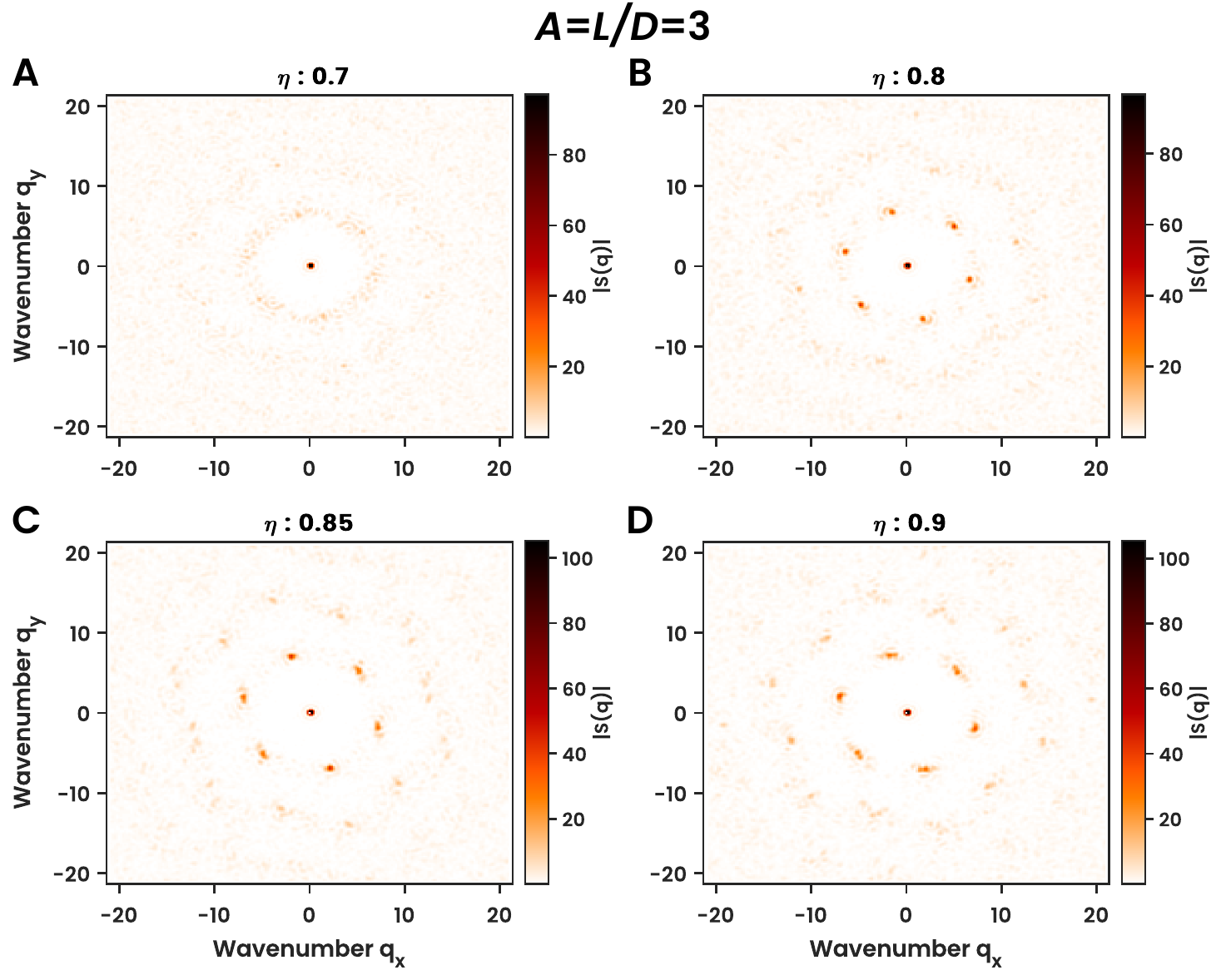}
    \caption{(A)-(D) Structure factor for different packing fractions $\eta$ for a system of SRS particles with shape anisotropy $A=3$ i.e. below the critical shape anisotropy $A_c$. It shows a gradual emergence of short-ranged positional ordering, only one or two rings in the structure factor plot.}
    \label{fig:structure_LD3}
\end{figure}

When the shape anisotropy is below the critical shape anisotropy $A_c$, we have seen that orientational defects are no longer possible in the range of packing fractions. Even within this range of packing fractions, there is a gradual emergence of short ranged hexagonal positional ordering. This can be seen in the case of $A=3$ (Fig. \ref{fig:structure_LD3}), which shows one or two rings in the structure factor plot at high packing fractions.

